\documentclass[superscriptaddress,reprint,amssymb,amsmath,aps,pra]{revtex4-1}
\usepackage{graphicx}
\usepackage{bm}
\usepackage{bbm}
\usepackage{amsthm}
\usepackage{braket}
\usepackage{amsfonts}
\usepackage{graphicx}

\begin{document}
\title{History state formalism for Dirac's theory} 
\author{N. L.\  Diaz}
\affiliation{Departamento de F\'isica-IFLP/CONICET,
	Universidad Nacional de La Plata, C.C. 67, La Plata (1900), Argentina}
\author{R. Rossignoli}
\affiliation{Departamento de F\'isica-IFLP/CONICET,
	Universidad Nacional de La Plata, C.C. 67, La Plata (1900), Argentina}
\affiliation{Comisi\'on de Investigaciones Cient\'{\i}ficas (CIC), La Plata (1900), Argentina}
\begin{abstract}
     We propose a history state formalism for a Dirac particle.
	 By introducing a reference quantum clock system  it is first shown that Dirac's equation can be derived by enforcing a timeless Wheeler-DeWitt-like equation for a global state. The Hilbert space of the whole system constitutes a unitary representation of the Lorentz group with respect to a properly defined invariant product, and the proper normalization of global states directly ensures standard Dirac's norm. Moreover, by introducing a second quantum clock, the previous invariant product emerges naturally from a generalized continuity equation.
	 The invariant parameter $\tau$ associated with this second clock labels history states for different particles, 
	 yielding an observable evolution in the case of an hypothetical  superposition of different masses. 
	 Analytical expressions for both space-time density and electron-time entanglement are provided for two particular families of electron's states, the former including 
     Pryce localized particles. 
\end{abstract}

\maketitle
\section{Introduction}

Time has been normally considered as an ``external'' parameter in quantum mechanics. In 1983 Page and Wootters \cite{W.83} introduced a formalism for non relativistic quantum mechanics where a reference quantum clock is introduced and the system evolution arises from an entangled system-clock history state satisfying a timeless Wheeler-DeWitt-like equation. Such formalism has recently received considerable attention and several extensions and consequences have been explored 
\cite{ML.15,Mo.14,Ma.15,FC.13,AB.16,Er.17,ML.17,Ni.18}.   
In this work our aim is to extend this approach to the relativistic regime, and specifically to a Dirac particle. It is first remarked that in this approach time operators act on the clock and not on the system, so that Pauli objection \cite{WP.80}  is circumvented. It is then shown that through a Wheeler-DeWitt like equation \cite{DW.67} for the global state, the Dirac equation \cite{PAM.28} naturally arises. The clock variable provides the time parameter of the equation. As a consequence, a difference between the present approach and the non relativistic case follows: the non absolute nature of time is introduced by defining the action of Lorentz transformations over global states. Lorentz symmetry is then preserved by introducing an invariant product in the complete Hilbert space.
The usual transformation of the wave function is then obtained, while the Hilbert space containing the global state provides a unitary representation of the proper Lorentz group. It is then shown that the appropriate normalization of free particle states under the 4-dimensional product, leads to the standard Dirac norm in ordinary 3-dimensional space in any frame of reference. These features allow a straightforward computation of expectation values of observables at a given time in a given frame of reference, completing the connection with the usual theory.

The addition of a second quantum clock with a second Wheeler-DeWitt-like equation enables to view the invariant density associated with the  previous product as that emerging in a generalized continuity equation. This addition follows the St\"uckelberg approach to relativistic quantum mechanics \cite{St.42}. 
 The time $\tau$ associated with this second clock labels history states for different particles, and while unobservable for fixed mass states, would lead to interference effects in a superposition of different history states.  
 It is also explicitly shown that for a time and mass independent potential, previous results remain valid, and entail a special orthogonality relation for degenerate eigenstates with different mass. 
 
 We finally discuss two features of the formalism: the space-time density induced by the invariant product, and the electron-time entanglement. For the former we prove that the density is positive definite in a family of solutions which include Pryce localized states \cite{Pr.48} in one spatial component. Moreover, we explicitly prove in the localized limit that it becomes null in the space-like region of the light cone with axes, say, $t$ and $z$, with $z$ the direction of localization. Furthermore, these properties extend to any mass distribution when the second clock is introduced.  We also provide general expressions for the eigenvalues of the reduced density matrix of the clock in the free particle case, which enables to evaluate the system-clock entanglement \cite{AB.16} in a given reference frame. These eigenvalues are frame dependent reflecting that in the present formalism both space and time are secondary variables \cite{PT.04,Te.05}. As an example, analytical expressions for any Lorentz frame are provided in a particular case.

\section{Formalism}
\subsection{Non relativistic case}
We first briefly review the Page-Wootters formalism \cite{W.83,ML.15}.  We set in what follows $\hbar=1$, $c=1$. 
Consider a bipartite system with Hilbert space $\mathcal{H}=\mathcal{H_{T}}\otimes \mathcal{H_{S}}$. The ``clock'' space $\mathcal{H_{T}}$ is spanned by the operator $T$ which satisfies the canonical commutation $[T,P_{T}]=i$. They whole system is assumed to be in a static pure state of the form
\begin{equation}\label{eq:clocksystemstate}
|\Psi\rangle=\int dt |t\rangle |\psi(t)\rangle\,.
\end{equation}
The state of the system is recovered by conditioning on the clock state:  $|\psi(t)\rangle=\langle t | \Psi \rangle$. 
Considering now states which satisfy the equation
\begin{equation}\label{eq:wdw}
\mathcal{J}|\Psi\rangle=0\,,
\end{equation}
with 
\begin{equation}\label{eq:wdwoperator}
\mathcal{J}=P_{T}\otimes \mathbbm{1}+\mathbbm{1}\otimes H\,,
\end{equation}
where $H$ is the hamiltonian of the system, unitary evolution is restored and the standard Schr\"odinger equation is recovered:
\begin{equation}
\langle t|\mathcal{J}|\Psi \rangle=0\Rightarrow i\frac{d}{dt}|\psi(t)\rangle=H|\psi(t)\rangle\,.
\end{equation}
This approach was recently examined in detail in \cite{ML.15}, where the implementation of measurements was also considered. The Pauli objection \cite{WP.80} of a time operator in quantum mechanics is circumvented: the operator acts on a different Hilbert space, and as a consequence it commutes with the system Hamiltonian \cite{ML.15,ML.17}. Moreover, the generators of space translations $\bm{P}_{S}$ commutes with the generator of time translations $P_T$, as it should, since space and time are independent degrees of freedom.

\subsection{Free Dirac's particle}
We now examine the relativistic extension of the previous scheme. The complete Hilbert space $\mathcal{H}_T\otimes \mathcal{H}_S$ constitutes a natural representation of Poincar\'{e} group when the space of the system $\mathcal{H}_S$ is $L^2(\mathbb{R}^3)$. On the other side, in order to discuss an electron (positron) theory, we set $\mathcal{H}_S=L^2(\mathbb{R}^3)\otimes \mathbb{C}^{4}$ in accordance to \cite{T.92}. An adequate choice of the inner product will preserve Lorentz symmetry.

A general state of the universe can be written as 
\begin{equation}
|\Psi\rangle=\sum_{\sigma=0}^{3}\int d^{4}p \;\Psi_{\sigma}(p)|p,\sigma \rangle\,,
\end{equation}
where $|p,\sigma \rangle=|p_{0}\rangle_{T}|\textbf{p},\sigma\rangle_{S}$ are the improper eigenstates of the operators $P_{\mu}$ (where for $\mu=0$ the operator acts on the clock space, while for $\mu=1,2,3$ it acts on the system space) and, say,  of $\sigma_{12}$ and $\gamma_0$ (here $\sigma_{\mu\nu}=\frac{i}{2}[\gamma_\mu,\gamma_\nu]$, with $\frac{\hbar}{2}\sigma_{\mu\nu}=\frac{\hbar}{2}\epsilon_{\mu\nu\rho}\Sigma_\rho$ the spin operator for $\mu,\nu=1,2,3$).  The states  $|p\rangle$, $|\sigma\rangle$ satisfy $\langle p'|p\rangle=\delta^{(4)}(p-p')$, 
$\langle \sigma|\sigma'\rangle=\delta_{\sigma \sigma'}$.
We introduce the adjoint system state $\langle\overline{\textbf{p},\sigma|}:= \langle \textbf{p}, \xi| \gamma^{0}_{\xi \sigma}$.
Because $d^{4}p$ is a Lorentz invariant measure we can introduce unitary boosts operators $U(\Lambda)$ in this space with respect to the product 
\begin{equation}\langle\bar{\Psi}_{1}|\Psi_{2}\rangle\equiv \int d^{4}p\; \bar{\Psi}_{1}(p)\Psi_{2}(p)\,,\label{6}
\end{equation}
where $\bar{\Psi}(p)=\Psi^{\dag}(p)\gamma^{0}$: 
\begin{equation}
U(\Lambda)|p,\sigma\rangle=S_{\sigma \xi}(\Lambda)|\Lambda p,\xi\rangle, \label{ust}\end{equation}
with $ \Lambda^{\mu}_{\;\nu}=e^{w^{\mu}_{\;\nu}}$ and $S(\Lambda)=e^{-\frac{i}{4}\sigma_{\mu \nu}w^{\mu \nu }}$  
\cite{ft1}. 
Unitarity follows from the property $S^{\dag}\gamma^{0}S=\gamma^{0}$ for time preserving Lorentz's transformations. The 
transformed state is then 
\begin{equation}
U(\Lambda) |\Psi \rangle=\sum_{\sigma=0}^{3} \int d^{4}p\, \Psi'_{\sigma}(p)|p,\sigma\rangle\,,
\end{equation}
with \begin{equation}\label{eq:transformation}  \Psi'_{\sigma}(p)= \langle p, \sigma|U(\Lambda)|\Psi \rangle=S_{\alpha \sigma} \Psi_{\alpha}(\Lambda^{-1}p)\,.
\end{equation}
 We may also define the states $|x,\sigma\rangle=|x^0\rangle|\textbf{x},\sigma\rangle=\frac{1}{(2\pi)^2}\int  d^4p\, e^{ipx}|p,\sigma\rangle$ with $px=p_\mu x^\mu$,  which, using Eq.\ (\ref{ust}), transform as  $U(\Lambda)|x,\sigma\rangle=S_{\sigma\xi}(\Lambda)|\Lambda x,\xi\rangle$. If the $|x,\sigma\rangle$ are the eigenstates of operators $X^{\mu}$, then the canonical commutation rules for both the clock and the system can be summarized as $\left[X^{\mu},P_{\nu}\right]=i \delta^{\mu}_{\;\nu}$.

The following step is to consider Eqs.\ (\ref{eq:wdw})-(\ref{eq:wdwoperator}) with $\mathcal{J}$ now constructed  with the free Dirac Hamiltonian $H_D=\bm{\alpha}\cdot\textbf{p}+\beta m$, 
\begin{equation}
{\cal J}=P_0\otimes \mathbbm{1}+\mathbbm{1}\otimes H_D\,.\label{JD}
\end{equation} Then ${\cal J}|\Psi\rangle=0$ leads to (setting $x^{0}=t$), 
\begin{equation}
\langle t|\mathcal{J}|\Psi \rangle=0\Rightarrow i\frac{d}{dt}|\psi(t)\rangle=(\bm{\alpha}\cdot\textbf{p}+\beta m)|\psi(t)\rangle\,,
\end{equation}
with $|\psi(t)\rangle=\langle t|\Psi\rangle=\frac{1}{\sqrt{2\pi}}\sum_{\sigma=0}^{3}\int d^{3}p\, e^{-ip_0 t}\Psi_{\sigma}(p)|\textbf{p},\sigma \rangle$. 
Equivalently, by defining $\mathbb{J}=-\gamma^\mu P_\mu$, we may rewrite Eq.\ (\ref{eq:wdw}) (an eigenvalue equation for ${\cal J}$ with eigenvalue 0) as an eigenvalue equation for $\mathbb{J}$ with eigenvalue $m$: 
\begin{equation}\label{eq:wdw2}
\gamma^{0}\mathcal{J}|\Psi\rangle=0 \Leftrightarrow -\gamma^{\mu}P_{\mu}|\Psi\rangle=m|\Psi\rangle\,.
\end{equation}
As a consequence of Pauli's fundamental theorem \cite{Sh.05},  $S^{-1}(\Lambda)\gamma^{\mu}S(\Lambda)=\Lambda^{\mu}_{\;\nu}\gamma^{\nu}$ and hence 
$U^{-1}(\Lambda)\gamma^\mu P_\mu U(\Lambda)=\gamma^\mu P_\mu$. Therefore, 
Eq.\ (\ref{eq:wdw2}) defines an invariant subspace, i.e.,
\begin{equation}
(\gamma^\mu P_\mu+m)|\Psi\rangle=0\Rightarrow (\gamma^\mu P_\mu+m)U(\Lambda)|\Psi\rangle=0\,.
\end{equation}
We can also rewrite Eq.\ (\ref{eq:wdw2}) in terms of $\Psi_{\sigma}(x):=\langle x,\sigma |\Psi\rangle$  recovering the covariant form of Dirac's equation \cite{PAM.28} (note that $\langle x,\sigma|P_\mu|\Psi\rangle=-i\partial_{\mu}\Psi_{\sigma}(x)$)

\begin{equation}\label{eq:dirac}
\langle x,\sigma |(\gamma^\mu P_\mu+m)| \Psi\rangle=0 \Rightarrow i\gamma_{\sigma \xi}^{\mu}\partial_\mu \Psi_\xi(x)=m\Psi_\sigma(x)\,.
\end{equation} 
States satisfying (\ref{eq:wdw2}) can be written in the form (in what follows sum over $\sigma$, $s$ and $r$ are implied) 
\begin{eqnarray}\label{eq:solutions}
|\Psi_m\rangle&=&
\int d^{4}p \; \delta(p^\mu p_\mu-m^2)H^+(p^0)u_{\textbf{p}\sigma}^{s}\, a_{s}(\textbf{p})|p,\sigma \rangle \nonumber
\\&&\oplus 
\int d^{4}p \; \delta(p^\mu p_\mu-m^2)H^-(p^0)v_{-\textbf{p}\sigma}^{r}\,b_{r}(\textbf{p})|p,\sigma \rangle\,,\nonumber\\
\end{eqnarray}
where, setting $E_{\textbf{p}}=\sqrt{\textbf{p}^2+m^2}$ 
\begin{subequations}
	\begin{align}
	u^{s}_{\textbf{p}\sigma}&=\frac{1}{\sqrt{E_{\textbf{p}}+m}}\begin{pmatrix}
	(E_{\textbf{p}}+m) \chi^{s} \\
	\textbf{p}.\boldsymbol{\sigma} \chi^{s} \\
	\end{pmatrix}_{\sigma} \,,\\
		v^{r}_{\textbf{p}\sigma}&=\frac{1}{\sqrt{E_{\textbf p}+m}}\begin{pmatrix}
	\textbf{p}.\boldsymbol{\sigma} \chi^{r} \\
		(E_{\textbf{p}}+m) \chi^{r} \\
	\end{pmatrix} _{\sigma}\,,
	\end{align}
\end{subequations}
with $s,r=0,1$.
The presence of the fourth ket implies orthogonality between particle and antiparticle subspaces for nonzero mass. We also notice that the sign in $|p^0\rangle$  implies a different sign in the evolution parameter $t$ between particle and antiparticle spaces after conditioning,  reflecting the Feynman-St\"uckelberg interpretation \cite{St.41,Fy.49}. 
In the subspace of solutions of Eq.\  (\ref{eq:wdw2}) the previous pseudoeuclidean inner product becomes isomorphic to two euclidean products. This is a consequence of the following relations \cite{Cel.16}:
\begin{subequations}
		\begin{eqnarray}
	\bar{u}^{r}_{\textbf{p}}u^{s}_{\textbf{p}}&=&2m \delta^{rs}\,,\\
	\bar{v}^{r}_{\textbf{p}}v^{s}_{\textbf{p}}&=&-2m \delta^{rs}\,.
	\end{eqnarray}
\end{subequations}

Since superposition of particles and antiparticles states are not realizable in nature \cite{Wg.74}, we will consider just one of the two terms of (\ref{eq:solutions}). In the following, we will work in the subspace of particles with positive mass. 
The overlap between states of different masses but same moment-spin distribution yields (see Appendix \ref{A} and \ref{B}) 
\begin{eqnarray}
\langle \bar{\Psi}_{m'}|\Psi_m\rangle&=&\int \frac{d^3p}{4E_{\textbf{p},m'}E_{\textbf{p},m}}\delta(E_{\textbf{p},m}-E_{\textbf{p},m'})\bar{u}^s_{\textbf{p},m}u^r_{\textbf{p},m'}\nonumber\\
&&\times a_s^*(\textbf{p},m)a_r(\textbf{p},m')\\
&=&\delta(m-m')\int \frac{d^3p}{2E_{\textbf{p}}} ||a(\textbf{p})||^2\label{eq:deltanorm}\,.
\end{eqnarray} 
The normalization $\langle \bar{\Psi}_m|\Psi_m\rangle=\delta(m-m')$ then implies  
$\int \frac{d^3p}{2E_{\textbf{p}}} ||a(\textbf{p})||^2=1$ and hence the Dirac norm (see below). 

An electron-clock state can be written  as (we omit the subscript $m$) 
\begin{eqnarray}
|\Psi\rangle&=&\frac{1}{\sqrt{2\pi}}\int d^{4}x\; {\psi}_{\sigma}(x)|x,\sigma\rangle\,, \\ {\psi}_{\sigma}(x)&=& \frac{1}{(2\pi)^{3/2}}\int \frac{d^{3}p}
{2 E_{\textbf p}}u_{\sigma}^s(\textbf{p})a_s(\textbf{p})
e^{-ipx|_{p^0=E_{\textbf p}}}\,.
\end{eqnarray}
From the invariance of $d^{4}x$ it follows the transformation law ${\psi}'_{\sigma}(x)=S_{\alpha \sigma} {\psi}_{\alpha}(\Lambda^{-1}x)$. Moreover, a simple calculation (see Appendix \ref{A}) shows that 
$\int d^3x\, \psi^\dagger (\textbf{x},t)\psi(\textbf{x},t)=\int \frac{d^3p}{2E_{\textbf{p}}} ||a(\textbf{p})||^2=1$, 
recovering the standard Dirac norm \cite{PAM.28}.  

The state of the electron can then be recovered by conditional probability as
\begin{equation}
|\psi(t)\rangle_{e}=\frac{\langle t|\Psi\rangle}{\sqrt{\langle \bar{\Psi}| \Pi^{\gamma_0}_{t}|\Psi\rangle}}
\end{equation}
with $\Pi^{\gamma_0}_{t}=|t\rangle\langle t|\otimes \gamma^{0}$ and  $\langle \bar{\Psi}| \Pi^{\gamma_0}_{t}|\Psi\rangle=\frac{1}{2\pi}\int d^3x\, \psi^\dagger (\textbf{x},t)\psi(\textbf{x},t)=\frac{1}{2\pi}$. 
The transformation law of the wave function implies the invariance of this quantity (see Appendix A). 
The correspondence with Dirac's theory is complete after noticing that the expectation value of an observable $M_e$, at a given time $t$, is obtained as follows:
\begin{equation}
\langle M_e \rangle (t)=\frac{\langle \bar{\Psi}|\Pi_{t}^{\gamma_0}M |\Psi\rangle}{\langle \bar{\Psi}|\Pi_{t}^{\gamma_0}| \Psi\rangle}= \ _e \langle \psi(t)|M_e|\psi(t)\rangle_e\,,
\end{equation}
where $M:=\mathbbm{1}\otimes M_{e}$.

As a final remark we write the general relation between the invariant product in 4-dimensional space with Dirac's product in ordinary 3-dimensional space with fixed mass $m$:
\begin{equation}
\langle \bar{\Phi}_{m'}|\Psi_m\rangle=\delta(m-m')\,(\phi,\psi)_m\,,
\end{equation}
where  we have defined $(\phi,\psi)_m:=\int d^{3}x\, \phi_m^{\dag}(\textbf{x},t)\psi_m(\textbf{x},t)$, while $\langle \bar{\Phi}_{m'}|\Psi_m\rangle=\frac{1}{2\pi}\int d^{4}x\, \bar{\phi}_{m'}(x)\psi_m(x)$.
\subsection{Bidimensional clock and proper time\label{C}}

We have seen that it is possible to enlarge the Hilbert space of the particle by including a clock,  preserving Lorentz's symmetry by defining an invariant product in this new space.  Moreover, for physical states satisfying a timeless equation, the notion of orthogonality which follows from this product, Eq.\ (\ref{eq:deltanorm}), yields the usual norm of Dirac's theory. In this section we will prove that the product we have introduced motivated by symmetry arguments arises naturally when a second clock is introduced. The aim is to discuss the usual identification of time in the Page-Wootters formalism with proper time \cite{ML.15}. 
While this identification is clearly satisfactory in the non relativistic case, the description of time evolution through Dirac's equation implies the necessity of introducing Lorentz transformations as nonlocal. This leads us to interpret the clock variable as time in a given reference frame. One may ask if there is a different approach to follow, in particular if it is possible to have  an equation analogue to (\ref{eq:wdw}) which after conditioning yields the evolution of the system state parametrized by an invariant variable $\tau$. Obtaining such an equation would mean to promote the role of $t$ to a dynamical variable, but this is exactly what the Page-Wootters formalism already does. It is not surprising then that by considering ``proper time'' in this way,  an extension of the formalism of the previous section ensues. We now develop this extension.

\subsubsection{Bidimensional clock}
Consider a bidimensional clock with Hilbert space $\mathcal{H}_{C}=L^2(\mathbb{R}^2)$ and basis $\{|\tau \rangle\otimes |t\rangle\}$, such that $\langle \tau'|\tau\rangle=\delta(\tau'-\tau)$ and $\langle t'|t\rangle=\delta(t-t')$, and the same Hilbert space $\mathcal{H}_S$ for the system as before.
A state of the whole system can be written as
\begin{equation}
|\Phi\rangle\rangle=\int d\tau \; |\tau\rangle |\Psi(\tau)\rangle=
\int dm \;\phi(m)|m\rangle |\Psi(m)\rangle\,,
\end{equation}
where $|\tau\rangle=\frac{1}{\sqrt{2\pi}}\int dm\, e^{-im\tau} |m\rangle$ and $|\Psi(\tau)\rangle=\frac{1}{\sqrt{2\pi}}\int dm\, \phi(m)e^{im\tau}|\Psi(m)\rangle \in \mathcal{H}_T\otimes \mathcal{H}_S$, the Hilbert space of the previous section.
We will assume that the Hamiltonian of the universe takes the form
\begin{equation}
\mathcal{J}= P_{\tau}\otimes\mathbbm{1}+\mathbbm{1}\otimes \gamma^{\mu}P_{\mu}\,.
\end{equation}
Notice that $\mathcal{J}$ has the same non-interacting form as before in the partition proper time--rest, but is non-separable in the partition clock--rest.   

Now, the equation 
\begin{equation}\mathcal{J}|\Phi\rangle\rangle=0\,,
\end{equation}implies $\langle \tau|{\cal J}|\Phi\rangle\rangle=0$, i.e., 
\begin{equation}\label{eq:stueckelberg}
i\partial_{\tau}|\Psi(\tau)\rangle=\gamma^{\mu}P_{\mu}|\Psi(\tau)\rangle\,,
\end{equation}
and, in the conjugate basis, 
\begin{equation}(\gamma^{\mu}P_{\mu}+m)|\Psi(m)\rangle=0\,.
\end{equation} This is the universe equation of the previous section, which determines an invariant subspace of $\mathcal{H}_T\otimes \mathcal{H}_S$ with respect to proper Lorentz transformations. This means that in the whole space $\tilde{U}(\Lambda):=\mathbbm{1}_\tau\otimes U(\Lambda)$ leaves the form of the equation (\ref{eq:stueckelberg}) invariant. In general, transformations leaving the form of (\ref{eq:stueckelberg}) invariant would also preserve it's square and hence a five dimensional metric, which  defines a Snyder space \cite{S.47}.  

By expanding the states $|\Psi(\tau)\rangle$ in the $|x,\sigma\rangle$ basis of $\mathcal{H}_T\otimes \mathcal{H}_S$ we obtain
\begin{subequations}\label{eq:stueckelbergwavefunction}	
	\begin{align}
		\gamma^{\mu}p_{\mu}\Psi(x,\tau)=i\partial_{\tau}\Psi(x,\tau)\,,\\
		\bar{\Psi}(x,\tau)\gamma^{\mu}p_{\mu}=i\partial_{\tau}\bar{\Psi}(x,\tau)\,,
	\end{align}
\end{subequations}
with $\Psi_{\sigma}(x,\tau):=\langle x,\sigma|\Psi(\tau)\rangle$ and $\bar{\Psi}(x,\tau):=\Psi^{\dag}(x,\tau)\gamma^0$. Therefore, 
\begin{equation}
\partial_{\mu}j^{\mu}(x,\tau)=-\frac{d}{d \tau} \bar{\Psi}(x,\tau)\Psi(x,\tau),
\end{equation} where $j^{\mu}(x,\tau):=\bar{\Psi}(x,\tau)\gamma^\mu \Psi(x,\tau)$, implying that for well behaved wavefunctions the quantity $\int d^4x\; \bar{\Psi}(x,\tau)\Psi(x,\tau)=\langle\bar{\Psi}(\tau)|\Psi(\tau)\rangle$ is conserved, i.e., the evolution operator $U(\tau)=e^{-i\gamma^{\mu}p_{\mu}\tau}$ preserves this norm. We see that the product we have chosen in the space $\mathcal{H}_T\otimes \mathcal{H}_S$ is the one which is preserved by $\tau$ evolution. Moreover
if we now expand in the mass basis and choose the normalization (\ref{eq:deltanorm}) we obtain
\begin{eqnarray}
 \langle\bar{\Psi}(\tau)|\Psi(\tau)\rangle&=&\!\!\int\!\! dm dm' \phi^{\ast}(m')\phi(m)e^{i\tau (m-m')}\langle\bar{\Psi}(m')|\Psi(m)\rangle\nonumber\\&=&\int dm\, |\phi(m)|^2\,.
\end{eqnarray}
Then we may choose $\int dm |\phi(m)|^2=1$ and interpret $\phi(m)$ as a mass distribution.

\subsubsection{The meaning of $\tau$ }
A scalar version of equation (\ref{eq:stueckelbergwavefunction}) with Hamiltonian $p^{\mu}p_{\mu}$ appeared several times in literature \cite{St.42,BM.00}, and a corresponding second order version was discussed in \cite{Fk.37}, where $\tau$ is identified with proper time.
In the present case, the classical (relativistic) momentum/speed relation for a free particle with proper time $\tau$ holds as an average computed with the induced product:
\begin{eqnarray}
\frac{d}{d\tau}\langle x^{\mu}\rangle&=&\int d^{4}x\; \bar{\Psi}(x,\tau) i\left[\gamma^{\nu}p_{\nu},x^{\mu}\right]  \Psi(x,\tau)\nonumber\\
&=&\int d^{4}x\;\bar{\Psi}(x,\tau) \gamma^{\mu}  \Psi(x,\tau)\nonumber\\
&=&\int\int \frac{dm d^{3}p}{2E_{\textbf{p},m}} |\phi(m)|^2||a(\textbf{p},m)||^2\left(\frac{p^{\mu}}{m}\right)=\langle \frac{p^{\mu}}{m}\rangle\,,
\nonumber\\
&&
\end{eqnarray} 
where he have used the Gordon identity \cite{Sh.05}.
Nevertheless, for a particle with definite mass, the $\tau$ evolution is trivial. As a consequence, the identification of $\tau$ with proper time is misleading. We may think instead that $\tau$ is parameterizing the relative phases of distinct particle's stories whose information is all condensed in the states $|\Psi(m)\rangle$ through the value of the mass and the moment-spin distribution. In a hypothetical superposition of different masses, i.e., different particles, it would become possible to see interference between separate stories and hence non trivial evolution in the parameter $\tau$.

\subsection{Dirac's particle in an external field}
A fully consistent description of interactions requires a field theory. Here we simply deal with the original Dirac's theory of a particle in an external field. We introduce the interaction by replacing ${\mathbb{J}}=-\gamma^{\mu}{P}_{\mu}$ by 
\begin{equation}\mathbb{J} _A=-\gamma^{\mu}({P}_{\mu}+e{A}_{\mu}(X))\,,\end{equation} 
with ${A}_{\mu}(X)|x\rangle=A_{\mu}(x)|x\rangle$. Then a state $|\Psi\rangle=\int d^{4}x\, \Psi_{\sigma}(x)|x,\sigma\rangle$ satisfies 
\begin{equation}
\mathbb{J}_A|\Psi\rangle=m|\Psi\rangle\,,
\end{equation} 
iff the wave function $\Psi(x)$ satisfies
\begin{equation}
\left(\gamma^{\mu}(-i\partial_{\mu}+eA_{\mu})+m\right)\Psi(x)=0\,.\label{Am}
\end{equation} 
We now focus on the case of a time independent $A^{\mu}$ in a given frame of reference. 
We first define the (normalized) eigenfunctions of  $H(m)=\bm{\alpha}\cdot(\textbf{p}+e\textbf{A})+\beta m+eA_0$, 
\begin{equation}\label{eq:eigenenergies}
H(m)\varphi_{kl}(\textbf{x},m)=E_k(m)\varphi_{kl}(\textbf{x},m)\,,
\end{equation}
where the subscript $l$ labels the eigenstates with the same energy. 
Then any solution of (\ref{Am}) is of the form  $\Psi(x)=\frac{1}{\sqrt{2\pi}}\sum_{k,l}c_{kl}e^{-iE_k(m)t}\varphi_{kl}(\textbf{x},m)$, which leads to
\begin{equation}\label{eq:state}
|\Psi_{m}\rangle=\sum_k c_k |E_k(m)\rangle|k(m)\rangle\,,
\end{equation} where  $c_k|k(m)\rangle=\sum_l c_{kl}\int d^{3}x\, \varphi^{\sigma}_{kl}(\textbf{x},m)|\textbf{x},\sigma\rangle$, with $|c_k|^2=\sum_l |c_{kl}|^2$ and 
 $\langle k'(m)|k(m)\rangle=\delta_{kk'}$, while 
$|E_k(m)\rangle=\frac{1}{\sqrt{2\pi}}\int dt \, e^{-iE_k(m)t}|t\rangle$. 

We now show that if potentials which depend on $m$ are excluded, e.g., \textit{gravity}, the condition $\langle \bar{\Psi}_{m'}|\Psi_m\rangle=\delta(m-m')$ {\it implies the usual normalization} $2\pi\int\,d^3x\,\Psi^\dag(\textbf{x},t)\Psi(\textbf{x},t)=\sum_k |c_k|^2=1$. 

\textit{Proof:} By using (\ref{eq:state}) we find, 
\begin{equation}\label{eq:normwithfield}
\langle\bar{\Psi}_{m'}|\Psi_m\rangle=\sum_{k, k'}c^*_{k'}c_{k}\delta(E_{k}(m)-E_{k'}(m'))\langle \overline{k'(m')}|k(m)\rangle\,.
\end{equation}
We will now prove the special orthogonality relation 
 \begin{equation}
\delta(E_{k}(m)-E_{k'}(m'))\langle \overline{k'(m')}|k(m)\rangle=\delta(m-m')\delta_{kk'}\,,\label{eq:iden}
\end{equation}
which implies  $\langle \bar{\Psi}_{m'}|\Psi_{m}\rangle=\delta(m-m')\sum_k |c_k|^2$,  where 
\begin{equation}\label{eq2:normwithfield}
\langle \overline{k'(m')}|k(m)\rangle=\sum_{l',l}\frac{c_{k'l'}^*c_{kl}}{c^*_{k'}c_k}\int d^{3}x\, \bar{\varphi}_{k'l'}(\textbf{x},m')\varphi_{kl}(\textbf{x},m)\,.
\end{equation}
We analyze the right hand side of  Eq.\ (\ref{eq2:normwithfield}) separately for $k=k'$ and $k\neq k'$.  

We first note that for $k=k'$ in (\ref{eq:iden}), $\delta(E_k(m)-E_{k}(m'))=\delta(m-m')/|dE_k(m)/dm|$. 
Deriving Eq.\ (\ref{eq:eigenenergies}) with respect to $m$ yields \[\left(H(m)-E_k(m) \right) \frac{d\varphi_{kl}(\textbf{x},m)}{dm}=\left(\frac{dE_k(m)}{dm}-\beta \right) \varphi_{kl}(\textbf{x},m)\,.
\]
Multiplying on the left by $\varphi^{\dag}_{kl'}(\textbf{x},m)$ and integrating over all space 
leads to the important result that these eigenfunctions satisfy the additional orthogonality condition 
\begin{equation}\int d^{3}x\, \bar{\varphi}_{kl'}(\textbf{x},m)\varphi_{kl}(\textbf{x},m)=\frac{dE_k(m)}{dm}\delta_{ll'}
\,, \end{equation}  
where we have used the hermiticity of $H(m)$ and the orthonormality of its eigenstates with respect to the usual product. The first part of the proof is complete assuming the standard result 
$dE_k(m)/dm>0$ for $E_k(m)>0$. 

The term with  $k\neq k'$ in (\ref{eq:iden})  contributes only when $E_{k'}(m')=E_{k}(m)$. 
Since 
\begin{eqnarray}
H(m)\,\varphi_{kl}(\textbf{x},m)&=&E_{k}(m)\varphi_{kl}(\textbf{x},m)\nonumber \\
H(m')\,\varphi_{k'l'}(\textbf{x},m')&=&E_{k'}(m')\varphi_{k'l'}(\textbf{x},m')\,, \nonumber
\end{eqnarray}
by multiplying on the left the first (second) equation by $\varphi^\dag_{k'l'}(m')$ ($\varphi^\dag_{kl}(m)$), integrating over all space and  subtracting the results (conjugating one of them), we find
\begin{eqnarray}
&&(m-m')\int d^{3}x\, \bar{\varphi}_{k'l'}(\textbf{x},m')\varphi_{kl}(\textbf{x},m)=\nonumber\\
&&[E_k(m)-E_{k'}(m')]
\int d^{3}x\, \varphi^\dagger_{k'l'}(\textbf{x},m')\varphi_{kl}(\textbf{x},m)\,.
\end{eqnarray}
Hence, if $E_k(m)=E_{k'}(m')$ the first integral should {\it vanish} for $m\neq m'$, 
implying that these eigenfunction satisfy in this case an extended orthogonality condition, which leads to the vanishing of (\ref{eq2:normwithfield}) for $k\neq k'$. 
 Note, however, that such orthogonality does not hold in general for $E_k(m)\neq E_{k'}(m')$. Previous results then lead to Eq.\ 
 (\ref{eq:iden}). \qed
   
It is then proved that whenever  a reference frame where   $A_\mu$  becomes  $t$ independent  exists, the invariant product implies Dirac's norm. We also mention that for $A^\mu$ independent of $t$ (and $\tau$) the extension of the treatment of section  (\ref{C}) is straightforward. 
 
\section{Invariant density and entanglement}
The formalism relies on the concept of the invariant product (\ref{6}) and the entanglement between the system and the reference clock. We now discuss some basic properties and examples. 
\subsection{The invariant density}
We now  examine in more detail the space-time density $\bar{\Psi}(x)\Psi(x)$ which corresponds to the invariant product $\langle \bar{\Psi}|\Psi\rangle$ we have introduced. Such density  is not positive-definite in either particle or antiparticle subspace. However, in the 1+1 dimensional case for the distribution $a(p)=e^{-\epsilon E_{p}}$ and a mass $m\neq 0$, it stays positive in all space-time. Moreover in the limit $\epsilon\to 0^{+}$ it becomes null in the space-like region of the light-cone centered in $(x,t)=(0,0)$. We also notice that the chosen distribution corresponds to the formal replacement $t\to t-i\epsilon$ in the case of a flat momentum distribution. Moreover, for $x\rightarrow z$, it can be regarded as a $3d$ distribution $\propto \delta(p_x)\delta(p_y)e^{-\epsilon E_{\textbf{p}}}$, in which case $\Psi(\textbf{x},t,\epsilon)$ becomes for $\epsilon\rightarrow 0^+$ and $t\to 0$ an eigenstate of the third component of the Pryce position operator $\textbf{q}=\textbf{x}+\frac{1}{2E_{\textbf{p}}^2}(\textbf{p}\times \bm{\Sigma}+im\beta\bm{\alpha})$ \cite{Pr.48}. 

Spinors in the 1+1 dimensional case have two components ($\sigma=0,1$) and fixed spin. The corresponding (unnormalized) wave function is (Eq.\ (\ref{eq:solutions})) 
\begin{equation}\label{eq:wavefunction1dim}
\psi_{\sigma}(x,t,\epsilon)=\int_{-\infty}^{\infty} \frac{dp}{2E_{p}}e^{-i(t-i\epsilon)E_{p}+ipx}	\frac{1}{\sqrt{E_{p}+m}}\begin{pmatrix}
E_{p}+m \\
p \\
\end{pmatrix}_\sigma\,,
\end{equation}
and satisfies the one dimensional equation $i\partial_t\psi(x,t)=-i\sigma_{1}\partial_x\psi(x,t)+m\sigma_{3}\psi(x,t)$. Now $\sigma_3$ replaces $\gamma^0$ when calculating $\bar{\Psi}(x,t)$. Thus, $\bar{\Psi}(x,t)\Psi(x,t)=\frac{1}{2\pi}[|\psi_0(x,t,\epsilon)|^2-|\psi_1(x,t,\epsilon)|^2]$. 

\begin{figure}[h]
	\begin{center}
\includegraphics[width=0.45\textwidth]{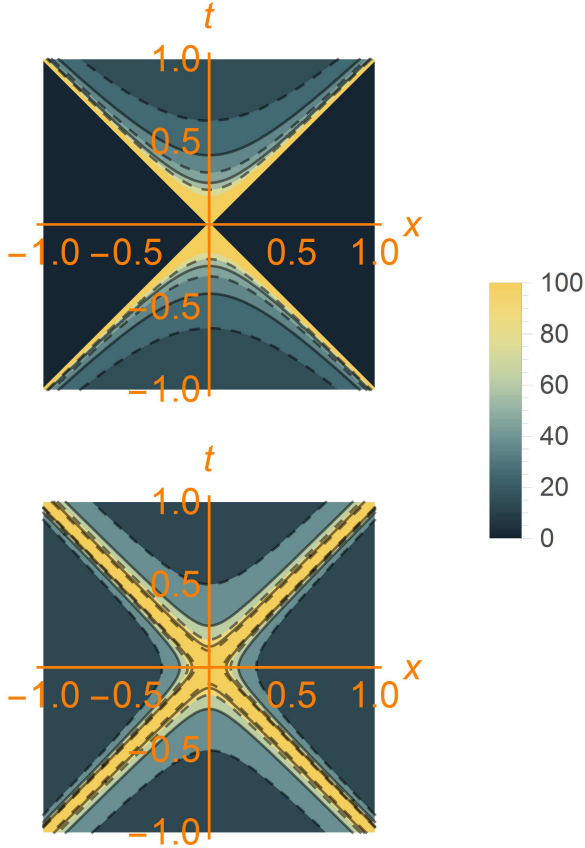}	\end{center}
	\caption{Contour plot of the invariant space time density (\ref{eq:density}) (top) and the Dirac density (\ref{eq:density2}) (bottom), for $m\equiv mc/\hbar=1$. The first one vanishes in the spacelike region (here $x$ and $t\equiv ct$ are in units of $\hbar/mc$).}\label{f1}
\end{figure}

By performing the integration in (\ref{eq:wavefunction1dim}) \cite{GR.07} it can be explicitly proved (see Appendix \ref{C}) that such difference is {\it positive} $\forall$ $x,t$ if $\epsilon>0$. And in the limit $\epsilon\rightarrow 0^+$, we obtain, for both $\bar{\psi}(x,t)\psi(x,t)$ and $\psi^{\dag}(x,t)\psi(x,t)$, 
\begin{eqnarray}\label{eq:density}
\bar{\psi}(x,t)\psi(x,t) &=& {\left\{
\begin{array}{cr}
\frac{\pi}{\sqrt{t^2-x^2}}&  x^2 < t^2\\
0 &  x^2 > t^2
\end{array}
\right.}\,,
\\
\psi^{\dag}(x,t)\psi(x,t) &=& \left\{
\begin{array}{lr}
\frac{\pi|t|}{t^2-x^2}&  x^2 < t^2\\
\frac{\pi|x|}{x^2-t^2}e^{-2m\sqrt{x^2-t^2}} &  x^2 > t^2
\end{array}
\right.\,.\label{eq:density2}
\end{eqnarray} 
Therefore, (\ref{eq:density}) is positive in the timelike sector,  vanishing in the spacelike region (see Fig.\ \ref{f1}). In contrast, (\ref{eq:density2}) stays positive in the latter \cite{T.92}. It is also  easy to show that 
$\lim\limits_{t\to0}\left(\lim\limits_{\epsilon\to0}\bar{\psi}(x,t,\epsilon)\psi(x,t,\epsilon)\right)\propto
\delta(x)$. 

In \cite{St.42} the Schr\"odinger-like density 
of the scalar version of Eq.\ (\ref{eq:stueckelbergwavefunction}) is interpreted as space-time probability density. In the present case the analogue quantity is given by $\bar{\Psi}(x,\tau)\Psi(x,\tau)\propto \int dm dm'\,\phi^{\ast}(m')\phi(m) e^{i(m-m')\tau}\bar{\psi}_{m'}(x,t)\psi_m(x,t)$. In the 1+1 dimensional case already discussed, and in the limit $\epsilon\to 0^{+}$ we find
\begin{equation}
\bar{\psi}_{m'}(x,t)\psi_m(x,t) = \left\{
\begin{array}{lr}
\frac{\pi}{\sqrt{t^2-x^2}}e^{-i(m-m')\sqrt{t^2-x^2}}&  x^2 < t^2\\
0 &  x^2 > t^2
\end{array}
\right.\,.
\end{equation}
As a consequence, $\bar{\Psi}(x,t,\tau)\Psi(x,t,\tau)$ vanishes outside the light cone for any mass distribution $\phi(m)$. Inside the light cone instead $\bar{\Psi}(x,t,\tau)\Psi(x,t,\tau)\propto \frac{1}{\sqrt{t^2-x^2}}|\Phi(\tau-\sqrt{t^2-x^2})|^2$, where $\Phi(\tau)$ indicates the Fourier transform of the function $\phi(m)$. We see that the positive region of the density, which corresponds to the inner part of the light-cone, stays positive under $\tau$ evolution, whereas the outer  part stays null. Moreover, in the general case $\epsilon>0$, $\bar{\Psi}(x,t,\tau)\Psi(x,t,\tau)>0$ for any mass distribution, as shown in Appendix \ref{C}. \\

\subsection{Electron-time entanglement}
It is clear from (\ref{eq:clocksystemstate}) that if there is no correlation between time and space-spin degrees of freedom the evolution is trivial.
In the work \cite{AB.16} the concept of \textit{system-time entanglement} was introduced as a measure of distinguishable quantum evolution, based on the entanglement between a system and the reference clock. In this section we apply these concepts to a particle with fixed mass, say an electron. 
In the following we adopt the convention that every trace over the spin has an additional $\gamma^{0}$ (i.e.,  the usual product is replaced by $\langle \sigma'|\sigma\rangle=\gamma^{0}_{\sigma' \sigma}$).
In order to quantify entanglement in the partition $\mathcal{H}_T\otimes \mathcal{H}_S$ we introduce the clock's reduced density matrix $\rho_T$ by tracing on space and spin degrees of freedom, we obtain:
\begin{subequations}
\label{ee}
\begin{eqnarray}
	 \rho_T&=&\frac{1}{2\pi}\int dtdt'd^{3}x\, \psi^{\dag}(\textbf{x},t')\psi(\textbf{x},t)|t\rangle\langle t'|\label{e1}\\&=&
	\int \frac{d^{3}p}{2E_{\textbf{p}}} ||a(\textbf{p})||^2|E_{\textbf{p}}\rangle \langle E_{\textbf{p}}|\nonumber\\
	 & \equiv &
	 \int dp\; \lambda^{2}(p)|p\rangle \langle p|\,,\label{e4}
\end{eqnarray}
\end{subequations}
where $|E_{\textbf{p}}\rangle=\frac{1}{\sqrt{2\pi}}\int e^{-iE_{\textbf{p} t}}|t\rangle dt$,  $|p\rangle:=\sqrt{\frac{dE(p)}{dp}}|E_{\textbf{p}}\rangle=\sqrt{\frac{p}{E(p)}}|E_{\textbf{p}}\rangle$ and $\lambda^2(p):=\frac{p}{2 }\int d\Omega\, ||a(\textbf{p})||^2$, the eigenvalues of $\rho_T$. 
It is now straightforward to compute entanglements measures based on entropies of the reduced density matrix of the clock.

\subsubsection{Different Lorentz observers}
Electron-time entanglement is not a Lorentz invariant quantity since boosts operators act non locally. 
In the present formalism space and time are both secondary variables \cite{PT.04}.
In order to  calculate the clock density in a boosted frame we first notice that (\ref{eq:transformation}) implies that the wave function which correspond to the new frame is \[\psi'_{\sigma}(x)=\frac{1}{(2\pi)^{3/2}}\int \frac{d^{3}p}{2E_{\textbf{p}}}S_{\alpha \sigma}(\Lambda)u^{s}_{\alpha}(\Lambda^{-1}\textbf{p})a_{s}(\Lambda^{-1}\textbf{p})e^{-ipx}\,,\] where we have used the invariance of both $p.x|_{p^{0}=E\textbf{p}}$ and the measure $\frac{d^{3}p}{2E_{\textbf{p}}}$. Then, using (\ref{e1}),  
\begin{equation}
\rho'_T=
\int \frac{d^{3}p}{(2E_{\textbf{p}})^2} a^{\ast}_{s'}(\Lambda^{-1}\textbf{p})a_s(\Lambda^{-1}\textbf{p})F^{s's}_{\Lambda}(\Lambda^{-1}\textbf{p})|E_{\textbf{p}}\rangle \langle E_{\textbf{p}}|\,,
\end{equation}
where $F^{s's}_{\Lambda}(\textbf{p})\equiv u^{s' \dag}_{ \textbf{p}}S^{\dag}(\Lambda)S(\Lambda)u^{s}_{\textbf{p}}$.
From the invariance of Dirac's normalization it follows that (see Appendix A) 
\begin{equation}
\begin{split}
F^{s's}_{\Lambda}(\Lambda^{-1}\textbf{p})=u^{s' \dag}_{\Lambda^{-1} \textbf{p}}S^{\dag}(\Lambda)S(\Lambda)u^{s}_{\Lambda^{-1} \textbf{p}}=\delta_{ss'} 2E_{\textbf{p}}\,.
\end{split}
\end{equation}
And finally the eigenvalues of $\rho'_{T}$ are simply
\begin{equation}\label{eq:boostedeigenvalue}
 \lambda^2(p,v)=\frac{p}{2}\int d\Omega\, ||a(\Lambda^{-1}\textbf{p})||^2. 
\end{equation}

\subsubsection{Proper frame}

As example, we now compute explicitly the relative entanglement, measured through the purity ratio 
\begin{equation}
R(v):=\frac{\int dp\, \lambda^4(p,v)}{\int dp\, \lambda^4(p,0)}\,,\label{pr}
\end{equation} 
for an electron with `proper' momentum distribution 
\begin{equation}||a(\textbf{p})||^2={\textstyle\frac{\epsilon}{4 \pi  m K_1\left(\epsilon m/2\right)}}e^{-\epsilon E_{\textbf{p}}/2}\label{md}\,,\end{equation}
where $K_1$ denotes the modified Bessel function, which admits an analytic evaluation. 
Eq.\  (\ref{eq:boostedeigenvalue}) allows us to calculate ${\rm Tr}(\rho_T^2)$ in every reference frame, with ($\gamma=1/\sqrt{1-v^2}$) 
\[||a(\Lambda^{-1}\textbf{p})||^2={\textstyle\frac{\epsilon}{4 \pi  m K_1\left(\epsilon m/2\right)}} e^{-\frac{\epsilon}{2}\gamma(E_{\textbf{p}}-vp\cos(\theta))}\,.\]
The result is, noting that $\lambda^2(p,v)=\frac{1}
{m\gamma vK_1(\epsilon m/2)}e^{-\epsilon\gamma E_{\textbf{p}}/2}\sinh(\epsilon\gamma vp/2)$ and using integrals of \cite{GR.07}, 
\begin{eqnarray}
R(v)=2\frac{\gamma K_1(\epsilon m)-K_1(\gamma \epsilon m)}{\gamma^2v^2\epsilon m K_2(\epsilon m)}\,.
\end{eqnarray}

\begin{figure}[h]
	\includegraphics[width=0.4\textwidth]{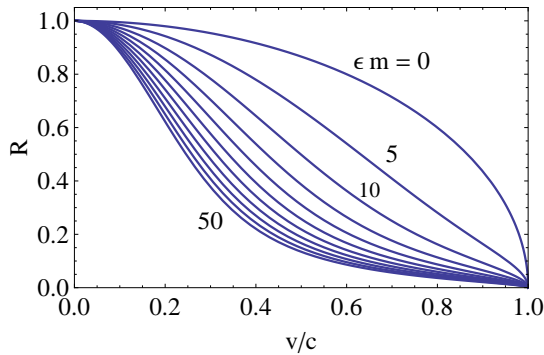}
		\caption{The purity ratio (\ref{pr}) for an electron with the proper distribution (\ref{md}) in terms of $v/c$ for different values of $\epsilon m$ (see text).}\label{f2}
	\end{figure}

This ratio is a decreasing monotonic function of  $v$, approaching $0$ for $v\rightarrow 1$, reflecting the increasing energy spread which leads to a vanishing purity ratio in this limit (see Fig.\ \ref{f2}). On the other hand, for $\epsilon m\rightarrow 0$,  $R(v)\rightarrow\gamma^{-1}=\sqrt{1-v^2}$ (and $\int dp\,\lambda^2(p,v)\rightarrow 1$). 
 
\section{Conclusions}
We have proposed a history state formalism for deriving Dirac's theory. 
The present approach enables to describe the theory within an explicit Hilbert space in a consistent manner, with an invariant norm which ensures the standard Dirac's norm for the wave function. The approach holds for a free particle as well as for a particle in a time (and mass)-independent external potential in a certain reference frame. In the presence of gravity, the formalism suggests, in principle,  that curvature effects may need to be considered even in the Newtonian limit. 

The inclusion of a second clock and an ensuing extended history state allowed us to derive the previous invariant norm precisely as that preserved by the evolution in the additional parameter. The latter would lead in principle to interference effects in a superposition of different mass states. We have also discussed some particular related aspects, like 
the invariant density and its positivity in the example considered,  which vanishes in the spacelike sector in the limit of a localized state (eigenstate of the Pryce position operator), in contrast with the Dirac density. We have also discussed the system-time entanglement for a free Dirac particle, obtaining analytic results for the purity ratio of the reduced clock density matrix according to different observers for a particular momentum distribution.  

The ideas developed in this work can be easily extended to  Klein-Gordon's theory. It could also constitute a suitable approach for a many particle theory, by considering a properly extended single particle space within a covariant field theory. These extensions are currently under investigation.

\acknowledgments
We acknowledge support from CIC, UNLP and CONICET of Argentina. 
Discussions with Prof.\ N.\ Gigena are also acknowledged. 
\appendix

\section{Dirac's norm \label{A}}

From the conservation of charge and the transformation law of the current $\bar{\psi}\gamma^{\mu}\psi$, it follows the invariance of the Dirac's norm \cite{BD.64}:
\begin{equation}\int d^{3}x\, \psi^{\dag}(\textbf{x},t)\psi(\textbf{x},t)=\int d^{3}x\, \psi'^{\dag}(\textbf{x},t)\psi'(\textbf{x},t)\,,\end{equation}
with $\psi'(\textbf{x},t)=S(\Lambda)\psi(\Lambda^{-1}x)$.
By expanding the wave function of a free particle in the momentum basis, and by using the property \cite{Cel.16} \[u^{(s')\dag}_{\textbf{p}}u^{(s)}_{\textbf{p}}=\delta^{ss'}2E_{\textbf{p}}\,,\]  we find 
\[\int d^{3}x\, \psi^{\dag}(\textbf{x},t)\psi(\textbf{x},t)=\int \frac{d^3p}{2E_{\textbf{p}}} ||a(\textbf{p})||^2\,,\] with $||a(\textbf{p})||^2:=|a_0(\textbf{p})|^2+|a_1(\textbf{p})|^2$.

From the invariance of both $p.x|_{p^{0}=E\textbf{p}}$ and the measure $\frac{d^{3}p}{2E_{\textbf{p}}}$, the equality (A1) can be restated as  
\[
\int \frac{d^3p}{2E_{\textbf{p}}} ||a(\textbf{p})||^2=\int \frac{d^3p}{2E_{\textbf{p}}2E_{\Lambda\textbf{p}}} a^{\ast}_{s'}(\textbf{p})a_{s}(\textbf{p})F^{s's}_{\Lambda}(\textbf{p})\,,
\]
implying the relation
\[
F^{s's}_{\Lambda}(\textbf{p})=u^{s' \dag}_{ \textbf{p}}S^{\dag}(\Lambda)S(\Lambda)u^{s}_{\textbf{p}}=\delta^{ss'} 2E_{\Lambda\textbf{p}}\,,
\]
where we have defined $F^{s's}_{\Lambda}(\textbf{p})=u^{s' \dag}_{ \textbf{p}}S^{\dag}(\Lambda)S(\Lambda)u^{s}_{\textbf{p}}$.

\section{State expansions in continuous variables \label{B}}
We consider a complete continuous set of states  $\{|p\rangle\}$ spanning a space ${\cal H}$ and satisfying 
$\langle p'|p\rangle=\delta(p-p')$,  and a state of the form 
\[|\psi\rangle=\int \phi(p)|p\rangle dp\,,\]
satisfying $\langle \psi|\psi\rangle=\int |\phi(p)|^2 dp=1$. 
If $E(p)$ is a monotonous function of $p$, we can rewrite $|\psi\rangle$ as 
\begin{eqnarray}|\psi\rangle&=&\int \phi(p(E))|p(E)\rangle \frac{dp}{dE} dE\label{B1}\\
&=&\int \Phi(E)|E\rangle dE\label{B2}\,,\end{eqnarray}
where $\Phi(E)=\phi(p(E))/\sqrt{|dE/dp|}$ and $|E\rangle=|p(E)\rangle/|\sqrt{dE/dp}|$, 
such that 
\[\int |\Phi(E)|^2 dE=1,\;\;\langle E'|E\rangle=\delta(E-E')\,.\]
The extension to states defined in ${\cal H}^{\otimes n}$ is apparent: the change from $n$ variables $p_i$  to new $n$ independent variables $E_i(\textbf{p})$ proceeds in the same way, with $|E_1\ldots E_n\rangle=|p_1\ldots,p_n\rangle/\sqrt{|J|}$ and 
$J$ the jacobian $\partial(E_1,\ldots,E_n)/\partial(p_1,\ldots,p_n)$. Note, however, that 
these states can be associated to different partitions of ${\cal H}^{\otimes n}$: 
If $P_i|p_i\rangle=p_i|p_i\rangle$, $[P_i,P_j]=0$, 
we may write $|p_1,\ldots,p_n\rangle=|p_1\rangle\ldots|p_n\rangle$ and similarly, 
$|E_1,\ldots,E_n\rangle=|E_1\rangle\ldots |E_n\rangle$, with  $H_i(\textbf{p})|E_i\rangle=E_i(\textbf{p})|E_i\rangle$ and  $[H_i,H_j]=0$. 

Considering now states in ${\cal H}\otimes {\cal H}$ of the form 
\[|\Psi\rangle=\int \phi(p,q)|pq\rangle dp dq\,,\]
we obviously have $\langle \Psi_1|\Psi_2\rangle=\int \bar{\phi}_1(p,q)\phi_2(p,q)dp dq$. 
And if $\phi_i(p,q)=g_i(p,q)\delta(f_i(p,q)-c_i)$, we obtain a finite overlap 
\begin{eqnarray}
\langle \Psi_1|\Psi_2\rangle&=&\!\!\int\!\! \bar{g}_1(p,q) g_2(p,q)\delta(f_1(p,q)-c_1)\delta(f_2(p,q)-c_2)dpdq\nonumber\\&=&
\bar{g}_1(p,q)g_2(p,q)/|J|\,,\end{eqnarray}
where $J=\partial (f_1,f_2)/\partial (p,q)$ and the final result is evaluated at the 
intersection of both curves (assumed here to exist and being unique; the extension to the general case is straightforward). On the other hand, if $f_1(p,q)=f_2(p,q)=f(p,q)$, 
we obtain, 
\begin{eqnarray}
\langle \Psi_1|\Psi_2\rangle&=&\!\!\int\! \bar{g}_1(p,q) g_2(p,q)\delta(f(p,q)-c_1)\delta(f(p,q)-c_2)dpdq\nonumber\\&=&
\delta(c_1-c_2)\int \bar{g}_1(p,q)g_2(p,q)dv/|J|\,, \end{eqnarray}
where the integral is along the curve $f(p,q)=c_1$, with $J=\partial(f,v)/(p,q)$ and 
$v(p,q)$ any function such that $(f,v)$ are independent variables.  
For instance $dv/|J|=dp/|f_q|$ if $v=p$.  Proper normalization of these states would then imply $\int \bar{g}_i(p,q)g_i(p,q)dv/|J|=1$. 

Note that these  states $|\Psi_i\rangle$ can  be written as 
\begin{eqnarray}|\Psi\rangle&=&\int g(p,q)\delta(f(p,q)-c)|pq\rangle dp dq\nonumber\\&=&\int g(p,q)|pq\rangle dv/|J|\\
&=&\int g(p,q)|pq\rangle dp/|f_q|\,,\end{eqnarray}
with the last two integrals over the curve $f(p,q)=c$, which defines the function $q(p)$ to be used in the last integral. Moreover, we can also rewrite the last integral in the more symmetric forms (using $|q(p)\rangle=|p\rangle/\sqrt{|dq/dp|}$, Eqs.\ (\ref{B1}--(\ref{B2})),
\begin{eqnarray}|\Psi\rangle&=&\int g(p,q)|p\rangle|p\rangle dp/\sqrt{|f_q f_p|} \nonumber\\&=&\int g(p,q)|q\rangle |q\rangle dq/\sqrt{|f_q f_p|}\,.\end{eqnarray}
These expressions represent continuous Schmidt decompositions of $|\Psi\rangle$. 

\section{Invariant Density \label{C}}
In order to prove that $\bar{\Psi}(x,t,\epsilon)\Psi(x,t,\epsilon)$ is positive for $\epsilon>0$, it sufficient to show that $F(x,t,\epsilon):=|\psi_{0}(x,t,\epsilon)/\psi_{1}(x,t,\epsilon)|^2> 1$. By performing the integration in (\ref{eq:wavefunction1dim}) \cite{GR.07} we find 
\begin{eqnarray}
\psi_0(x,t,\epsilon)&=&{\frac{\sqrt{2\pi}\sqrt{\sqrt{x^2-(t-i\epsilon)^2}+i(t-i\epsilon)}e^{-m\sqrt{x^2-(t-i\epsilon)^2}}}
	{2\sqrt{x^2-(t-i\epsilon)^2}}}\,,\nonumber\\&&\\
\psi_1(x,t,\epsilon)&=&\frac{\sqrt{2\pi}ixe^{-m\sqrt{x^2-(t-i\epsilon)^2}}}
{2\sqrt{x^2-(t-i\epsilon)^2}\sqrt{\sqrt{x^2-(t-i\epsilon)^2}+i(t-i\epsilon)}}\,,\nonumber\\&&
\end{eqnarray}
and hence,
\begin{eqnarray}\label{eq:F}
F(x,t,\epsilon)&=&1+\nonumber\\
&&{\textstyle\frac{2\sqrt{f(x,t,\epsilon)} \left(t \sin \left(\frac{\gamma}{2} \right)+\epsilon \cos \left(\frac{\gamma}{2}
	\right)\right)+f(x,t,\epsilon)-(x^2-\epsilon^2-t^2)}{x^2}}\,,\nonumber\\
&&
\end{eqnarray}
where $f(x,t,\epsilon)=\sqrt{\left(x^2-\epsilon^2-t^2\right)^2+4 x^2\epsilon^2}$ and
$\gamma(x,t,\epsilon):=\arg(x^2+\epsilon^2-t^2+2i\epsilon t)$. Notice that $F(x,t,\epsilon)$ is independent of $m$. For $\epsilon>0$ and  $t\geq 0$,   $0\leq\gamma\leq\pi$ while for $t\leq0$, $-\pi\leq\gamma\leq0$. In both cases $t \sin \left(\frac{\gamma}{2} \right)\geq0, \;\cos \left(\frac{\gamma}{2}\right)\geq0$. Then the quotient in (\ref{eq:F}) is clearly positive.  On the other hand, for $\epsilon=0$, 
$\gamma=0$ and the quotient becomes $(|x^2-t^2|-(x^2-t^2))/x^2$, 
implying $F(x,t,0)=1$ if $|x|>|t|$ and $F(x,t,0)=2t^2/x^2-1$ if $|x|<|t|$. 
From Eq.\ (\ref{eq:wavefunction1dim}) we notice, by performing the integral,  that  $\psi_{0}^{\ast}(x,t,\epsilon,m')\psi_0(x,t,\epsilon,m)-\psi_{1}^{\ast}(x,t,\epsilon,m')\psi_1(x,t,\epsilon,m)=\psi_{1}^{\ast}(x,t,\epsilon,m')\psi_1(x,t,\epsilon,m)\left(F(x,t,\epsilon)-1\right)$, with $F(x,t,\epsilon)$ 
defined in (\ref{eq:F}). This implies $\bar{\Psi}(x,t,\tau)\Psi(x,t,\tau)\propto |\int dm\, \phi(m)e^{im\tau }\psi_1(x,t,\epsilon,m)|^2\left(F(x,t,\epsilon)-1\right)>0$ since $F(x,t,\epsilon)>1$.


\begin{thebibliography}{999}
\bibitem{ft1}The unbounded continuous spectrum of the clock ensures that the state  (\ref{ust}) 
	is well defined. The active representation of the boost 
	preserves the partition. This guarantees that any observer is able to identify the clock.
\end{thebibliography}
\end{document}